\documentstyle[aps,prd,eqsecnum]{revtex}

\def\E{{\cal E}}

\def\R{{\cal R}}
\def\K5{\kappa_5}
\def\dqE{{\delta q_\E}}

\newcommand{\square}{\kern1pt\vbox{\hrule height  1.2pt\hbox{\vrule
width 1.2pt\hskip 3pt
\vbox{\vskip 6pt}\hskip  3pt\vrule width 0.6pt}\hrule height
0.6pt}\kern1pt}

\begin{document}

\preprint{hep-th/0012044}
\draft

% UNCOMMENT FOR TWO-COLUMN MODE
% \twocolumn[\hsize\textwidth\columnwidth\hsize\csname
% @twocolumnfalse\endcsname

\title{Large-scale cosmological perturbations on the brane}

\author{David Langlois$^1$, Roy Maartens$^2$, Misao Sasaki$^3$
and David Wands$^2$}
\address{~}
\address{$^1$Institut d'Astrophysique de Paris,
98bis Boulevard Arago, 75014~Paris, France}
\address{$^2$Relativity and Cosmology Group,
School of Computer Science and Mathematics,\\ University of
Portsmouth, Portsmouth~PO1~2EG, Britain}
\address{$^3$Department of Earth and Space Science,
Graduate School of Science,\\
Osaka University, Toyonaka~560-0043, Japan}
\date{\today}

\maketitle

\begin{abstract}
  In brane-world cosmologies of Randall-Sundrum type, we show that
  evolution of large-scale curvature perturbations may be determined
  on the brane, without solving the bulk perturbation equations.  The
  influence of the bulk gravitational field on the brane is felt
  through a projected Weyl tensor which behaves effectively like an
  imperfect radiation fluid with anisotropic stress.  We define
  curvature perturbations on uniform density surfaces for both the
  matter and Weyl fluids, and show that their evolution on large
  scales follows directly from the energy conservation equations for
  each fluid.  The total curvature perturbation is not necessarily
  constant for adiabatic matter perturbations, but can change due to
  the Weyl entropy perturbation.  To relate this curvature
  perturbation to the longitudinal gauge metric potentials requires
  knowledge of the Weyl anisotropic stress which is not determined by
  the equations on the brane.  We discuss the implications for
  large-angle anisotropies on the cosmic microwave background sky.

\end{abstract}

\pacs{04.50.+h, 98.80.Cq \hfill PU-RCG-00/37, OU-TAP-151,
hep-th/0012044}

% UNCOMMENT FOR TWO-COLUMN MODE
% \vskip2pc]

\section{Introduction}

Recently, a lot of attention has been devoted to the cosmology of
a brane-universe embedded in a higher dimensional spacetime,
stimulated by suggestions from string theory that there may exist
extra dimensions which are large but inaccessible to ordinary
matter~\cite{Antoniadis}.
In this paper we will investigate the simplest such model
with a single extra dimension, obeying the five-dimensional
Einstein equations in the bulk, with matter fields confined to a
single brane located at a $Z_2$-symmetric fixed point.

The standard Friedmann cosmology is not recovered in such a
model~\cite{BDL}, unless one assumes the existence of a constant
tension in the brane~\cite{etc,BDEL}
(in addition to ordinary matter) and a suitably adjusted negative
cosmological constant in the bulk, as in the (second)
Randall-Sundrum model~\cite{RS2}. The evolution is then
indistinguishable from the standard one in the low-energy regime
where the matter density in the universe is much smaller than the
brane tension. 
Therefore the background brane cosmology reproduces the
properties of the standard Friedmann background cosmology at the
present day if one requires standard evolution since at least
nucleosynthesis.
To be able to discriminate between a brane cosmology and standard
four-dimensional cosmology,
it is necessary to go one step further and study perturbations about
the background models.  Present and future data on large-scale
structure and cosmic microwave background (CMB) anisotropies provide
extensive information on the spectrum and evolution of cosmological
perturbations.

Our purpose here is to present the evolution equations for
perturbations in brane cosmology as close as possible to the standard
four-dimensional approach in order to discuss the possible imprint of
the fifth dimension on cosmological observations, and in particular
CMB anisotropies.  The influence of the bulk gravitational field on
the brane is felt through a projected Weyl tensor which behaves
effectively like an imperfect radiation fluid with anisotropic stress.
The present work results from the combination of two approaches to
brane perturbations: a covariant approach~\cite{RoyI,RoyII} based on
the effective four-dimensional Einstein equations on the
brane~\cite{SMS}, and a metric-based approach treating the bulk metric
perturbations in a Gaussian normal coordinate
system~\cite{Langlois,L2} (see also~\cite{GS,Mukplus,DDK} for other
metric-based approaches, and~\cite{GenSas} for a covariant Green's
function approach). We will also follow the approach adopted in
Refs.~\cite{chaotic,WMLL} in using the energy conservation equations
to compute the evolution of large-scale curvature perturbations and to
identify the effect of non-adiabatic modes.

We focus our attention on the evolution of large-scale perturbations,
i.e., perturbations on scales larger than the Hubble radius.  The
reason is that the scales of cosmological interest (e.g., for
large-angle CMB anisotropies) have spent most of their time far
outside the Hubble radius and have re-entered only relatively recently
in the history of the Universe.  Large-scale perturbations generated
from quantum fluctuations during de Sitter inflation on the brane have
been calculated~\cite{chaotic,HHR,LMW,bmw}. The spectrum of tensor
perturbations contains a massless zero mode and massive modes that
remain in the vacuum state~\cite{GS,LMW}. The amplitude of the zero
mode is enhanced at high energies compared with the usual
four-dimensional result, and is constant on large scales~\cite{LMW}.
In~\cite{bmw}, the vector metric perturbations were shown to have no
normalizable zero mode, while the normalizable massive modes remain in
the vacuum state during inflation (see also \cite{GenSas}). It follows
that large-scale vector and tensor perturbations from brane inflation
are expected to have the same qualitative properties as in general
relativity (apart from an enhanced tensor amplitude), and therefore
the same qualitative impact on CMB anisotropies.

Scalar perturbations were computed in~\cite{chaotic}, and it was shown
that the amplitude of the curvature perturbation on uniform density
hypersurfaces is enhanced at high energies relative to the standard
four-dimensional result (see also~\cite{cll,L2}).  In~\cite{chaotic},
the effects of the bulk Weyl tensor on the brane were neglected.
Large-scale scalar perturbations, incorporating the full bulk Weyl
effects, have
been investigated via a comoving covariant approach
in~\cite{RoyI}, where it was shown that, even when bulk Weyl
effects are included, the covariant density perturbation equations
contain a closed system on the brane without solving the bulk
perturbation equations. In~\cite{RoyII}, it was then shown that
the covariant analog of the longitudinal gauge metric potential
due to matter perturbations is non-constant on large scales in the
early universe.

In this paper, we define the curvature perturbation on uniform density
surfaces for matter and an entropy perturbation due the `Weyl' fluid.
Their evolution on large scales follows directly from the energy
conservation equations for each fluid.  The total curvature
perturbation is not necessarily constant for adiabatic matter
perturbations, but can change due to the Weyl entropy perturbation.
We go further to show that, while our approach is sufficient to
determine the curvature perturbation at late times due to matter and
Weyl effective density perturbations, it cannot determine the
anisotropic stress exerted on the brane by the projected Weyl tensor,
and hence the contribution of the scalar shear to CMB anisotropies.
Thus the effect of brane-world scalar perturbations on large-angle CMB
anisotropies cannot in general be determined in the same simple way
used in general relativity. Further investigation is required to solve
the bulk perturbation equations and determine the behavior of the Weyl
anisotropic stress.

\section{Field equations}

\subsection{Five-dimensional equations}

We assume that the gravitational field in the bulk obeys the
five-dimensional Einstein equations
\begin{equation}
\label{5DEinstein}
{}^{(5)\!}G_{AB} + \Lambda_5 \, {}^{(5)\!}g_{AB}
=\kappa_5^2\,{}^{(5)\!}T_{AB}\,,
\end{equation}
where
$\kappa_5^2$ is the five-dimensional gravitational constant and
$\Lambda_5$ the cosmological constant in the bulk. We further
assume that the spacetime is vacuum except at the brane.
The gravitational
field is also subject to appropriate boundary conditions at the brane.
The energy-momentum tensor for matter on the brane, $T_{\mu\nu}$,
and the brane tension, $\lambda$,
cause a discontinuity in the extrinsic curvature,
$K_{\mu\nu}$, given
by the junction conditions~\cite{BDL,SMS}
\begin{equation}
\label{junction}
\left[ K_{\mu\nu} \right]^+_-= - \kappa_5^2
 \left\{ \frac{1}{3}(\lambda-T) g_{\mu\nu}
+ T_{\mu\nu}\right\} \,,
\end{equation}
where $T=g^{\mu\nu}T_{\mu\nu}$,
$~K_{\mu\nu}=g_\mu^Ag_\nu^B\,^{(5)\!}\nabla_An_B$,
$~n^A$ is the spacelike
unit normal to the brane, and the projected metric on the brane
is given by
\begin{equation}
\label{gmunu}
g_{AB}={}^{(5)\!}g_{AB}-n_A n_B \,.
\end{equation}
We note that the division of the energy-momentum tensor into
$T_{\mu\nu}$ and $\lambda g_{\mu\nu}$ is rather arbitrary;
we choose $\lambda$ in such a way that the original
Randall-Sundrum
brane is recovered when $T_{\mu\nu}=0$.
If we assume that the brane is located at a $Z_2$-symmetric orbifold
fixed point, then the matter energy-momentum tensor
and the brane tension determine the
extrinsic curvature close to the brane:
\begin{equation}
\label{Kmunu}
K_{\mu\nu} = -\frac{\kappa_5^2}{2}
 \left[ \frac{1}{3}(\lambda-T) g_{\mu\nu}
+ T_{\mu\nu}\right] \,.
\end{equation}

\subsection{The view from the brane}

The effective
4-dimensional Einstein equations on the brane can be
obtained~\cite{SMS} by projecting the 5-dimensional quantities.
The Gauss equation leads to the 4-dimensional effective equations:
\begin{equation}
G_{\mu\nu}
 = - {\Lambda_5\over2}g_{\mu\nu}
+ KK_{\mu\nu}
-K_\mu{}^{\sigma}K_{\nu\sigma} -{1 \over 2}g_{\mu\nu}
  \left(K^2-K^{\alpha\beta}K_{\alpha\beta}\right) - \E_{\mu\nu}\,,
\label{4dEinstein}
\end{equation}
where
$K=K^\mu{}_\mu$ and the effect of the non-local bulk
gravitational field is described by the projected 5-dimensional
Weyl tensor
\begin{equation}
\E_{\mu\nu} \equiv {}^{(5)\!}C^E{}_{A F B}\,n_E n^F
g_\mu{}^{A} g_\nu{}^{B} \,.
\label{Edef}
\end{equation}
Using the junction conditions given in Eq.~(\ref{Kmunu}), we can
give the extrinsic curvature in terms of the energy-momentum tensor on
the brane so that
\begin{eqnarray}
\label{modEinstein}
G_{\mu\nu}+\Lambda_4 g_{\mu\nu}
= \kappa_4^2 T_{\mu\nu}+\kappa_5^4\,\Pi_{\mu\nu}
-\E_{\mu\nu}\,, \label{eq:effective}
\end{eqnarray}
%
%==========================%
where
\begin{eqnarray}
\Lambda_4 &=&{\Lambda_5\over2}
+{\kappa_5^4\over12}\,\lambda^2  \,,
\label{Lamda4}\\
\kappa_4^2\ &=& 8\pi G_{\rm N}={\kappa_5^4\,\over 6}\lambda\,,
\label{GNdef}\\
\Pi_{\mu\nu}&=&
-\frac{1}{4} T_{\mu\alpha}T_\nu{}^{\alpha}
+\frac{1}{12}TT_{\mu\nu}
+\frac{1}{24}\left(3T_{\alpha\beta}T^{\alpha\beta}-T^2\right)
g_{\mu\nu}\,.
\label{pidef}
\end{eqnarray}
Using the arbitrariness in the choice of $\lambda$ as noted before,
we set $\Lambda_4=0$.
The usual conservation laws for matter, $\nabla_\mu
T^\mu_\nu=0$, still apply (they are obtained by substituting
Eq.~(\ref{Kmunu})
into the Codazzi equations~\cite{SMS}).

The power of this approach is that the above form of the
4-dimensional effective equations of motion is independent of the
evolution of the bulk spacetime, being given entirely in terms of
quantities defined on the brane. Thus these equations apply to
brane-world scenarios with infinite or finite bulk, stabilised or
evolving.

Near the brane it is always possible to use Gaussian normal
coordinates $x^A=(x^\mu,y)$
in which the 5-dimensional line-element takes the form
\begin{equation}
^{(5)\!}ds^2 = g_{\mu\nu}(x^\alpha,y) dx^\mu dx^\nu + dy^2 \,,
\end{equation}
where the brane is located at $y=0$.

\section{Cosmological perturbations on the brane}

The most general linear scalar metric perturbation about
a Friedmann-Robertson-Walker (FRW) brane is~\cite{KS}
\begin{equation}
\label{pertmetric}
g_{\mu\nu} = \left[
\begin{array}{ccc}
-(1+2A) & & aB_{|i} \\ && \\
aB_{|j} & & a^2 \left\{ (1+2\R) \gamma_{ij} + 2E_{|ij} \right\}
\end{array}
\right] \,,
\end{equation}
where $a(t)$ is the scale factor, $\gamma_{ij}$ is the metric
for a maximally symmetric 3-space with comoving curvature $K=0,\pm1$,
and a vertical bar denotes the covariant derivative of $\gamma_{ij}$.

The perturbed energy-momentum tensor for matter on the brane,
with background energy density $\rho$ and pressure $P$, can be
given as
\begin{equation}
\label{Tmunu}
T^\mu_\nu =
\left[
\begin{array}{ccc}
-(\rho+\delta\rho) & & a(\rho+P) ( v +B )_{|j} \\ &&\\
-a^{-1}(\rho+P)v^{|i} & & (P+\delta P)\delta^i_j + \delta\pi^i_j
\end{array}
\right] \,,
\end{equation}
where
$\delta\pi^i_j=\delta\pi^{|i}{}_{|j}-{1\over3}
\delta^i_j\delta\pi^{|k}{}_{|k}$
is the tracefree anisotropic stress perturbation.
The perturbed quadratic energy-momentum tensor is
(compare~\cite{RoyI})
\begin{equation}
\Pi^\mu_\nu =
{\rho\over12}
\left[
\begin{array}{ccc}
-(\rho+2\delta\rho) & & 2a(\rho+P)(v +B )_{|j} \\&&\\
-2a^{-1}(\rho+P)v^{|i} & &
\left\{2P+\rho+2(1+P/\rho)\delta\rho + 2\delta P\right\}\delta^i_j
- (1+3P/\rho)\delta\pi^i_j
\end{array}
\right] \,.
\end{equation}
The remaining contribution of metric perturbations in the bulk to the
modified Einstein equations on the brane is given by the projected
Weyl tensor $\E^\mu_\nu$.  Although this is
due to the effect of bulk metric
perturbations not defined on the brane, we can nonetheless
parametrize this as an effective energy-momentum tensor~\cite{RoyI,L2}
\begin{equation}
-\E^\mu_\nu
 = \kappa_4^2
\left[
\begin{array}{ccc}
-(\rho_\E+\delta\rho_\E) & & a\dqE_{|j} \\&&\\
-a^{-1}\dqE^{|i}+a^{-1}(\rho_\E+P_\E)B^{|i} & &
(P_\E+\delta P_\E)\delta^i_j
 + \delta\pi_{\E}{}^{i}{}{}_{j}
\end{array}
\right] \,.
\end{equation}

In the background FRW cosmology,
Eq.~(\ref{modEinstein}) yields the modified
Friedmann equation
\begin{equation}
\label{modFriedmann}
3H^2 + {3K \over a^2} = \kappa_4^2 \rho \left( 1+{\rho\over2\lambda}
\right) + \E_0^0 \,,
\end{equation}
where $H=\dot{a}/a$ is the Hubble expansion rate.
For matter on the brane, one can
define an
effective gravitational energy density and pressure
\begin{eqnarray}
\rho_{\rm eff} &=& \rho \left( 1+{\rho\over2\lambda} \right)
\,, \label{rhoeff}\\
P_{\rm eff} &=& P + {\rho \over 2\lambda} \left( 2P + \rho \right) \,,
\label{Peff}
\end{eqnarray}
which obey the standard Friedmann equation
if $\E^0_0=\kappa_4^2\rho_{\E}$ vanishes.
The  adiabatic sound speed for matter is given by
$c_{\rm s}^2=\dot{P}/\dot\rho$, and
the effective adiabatic sound speed for the effective matter fluid is
given by
\begin{equation}
c_{\rm eff}^2 = {\dot{P}_{\rm eff} \over \dot{\rho}_{\rm eff}}
 = c_{\rm s}^2 + \left( {\rho+P\over \rho+\lambda} \right) \,.
\end{equation}

In the low energy regime, $\rho\ll \lambda$, the effective density
and pressure tend to the real quantities, and
standard cosmology is recovered, up to
the Weyl term, as is clear from Eq.~(\ref{modFriedmann}).
Given the agreement between the abundance of light elements and the
nucleosynthesis predictions, there is not much freedom for
non-standard evolution
of the scale factor from the time of nucleosynthesis.
The Universe is thus in a low-energy regime since at least
nucleosynthesis, i.e. $\lambda > \rho_{\rm nucl}\sim (1~{\rm MeV})^4$.
This implies a lower bound on the 5-dimensional mass $M_5$ (defined by
$\K5^2=8\pi/M_5^3$) of~\cite{etc}
$M_5\gtrsim 10$ TeV.
In fact, it turns out there is a more stringent
constraint coming from small-scale gravity experiments; the absence
of deviations from Newton's law on the millimeter scale (see,
e.g.~\cite{newton}) imposes~\cite{chaotic} $M_5\gtrsim 10^5$ TeV,
corresponding to
\begin{equation}
\lambda^{1/4} \gtrsim 100 \ {\rm GeV} \, .
\end{equation}
At times much earlier than nucleosynthesis, there is no a priori
argument against a non-standard evolution of the Universe. In a
very high-energy regime, $\rho\gg \lambda$, the contribution from
the matter in Eq.~(\ref{modFriedmann}) becomes quadratic in the
energy density. Thus a barotropic fluid with
$P/\rho=c_s^2=$constant, leads to an effective sound speed given
by $c_{\rm eff}^2=2c_s^2+1$ at high energies ($\rho\gg\lambda$).
In particular, ordinary radiation with $c_s^2={1\over3}$ yields an
effective sound speed given by $c_{\rm eff}^2={5\over3}$ at high
energies in the early brane-world universe.

There is an additional contribution to the modified Friedmann
equation~(\ref{modFriedmann}) from the projected Weyl tensor,
equivalent to an additional energy density,
$\rho_\E=\E_0^0/\kappa_4^2$. The tracefree property of
$\E^\mu_\nu$ implies that the pressure obeys
$P_\E={1\over3}\rho_\E$ and the effective sound speed is given by
$c_\E^2={1\over3}$.

We define a total effective energy density and pressure on the brane:
\begin{eqnarray}
\rho_{\rm tot} &=& \rho \left( 1+{\rho\over2\lambda} \right) + \rho_\E
\,,\\
P_{\rm tot} &=& P + {\rho\over2\lambda} \left( 2P+\rho \right)
+{1\over 3}\rho_\E \,.
\end{eqnarray}
There are again constraints on the contribution of $\rho_\E$ to the
total energy density in the Universe~\cite{BDEL} from nucleosynthesis;
since the effective number of light neutrino species must be less than
$3.2$~\cite{BKT} we have
\begin{equation}
\label{nuclimit}
\frac{\rho_\E}{\rho} \lesssim 0.03
\end{equation}
at the time of nucleosynthesis. This implies, since
$\rho_\E\propto 1/a^4$, that the Weyl contribution would be
extremely small today.

\subsection*{Conservation equations}

Together with the junction conditions at the brane, the
4-dimensional modified Einstein equations~(\ref{eq:effective}) are
equivalent to the 4-dimensional part of the 5-dimensional
Einstein equations.
Two of the remaining 5-dimensional Einstein equations are equivalent
to the conservation of the matter energy and momentum on the brane:
\begin{eqnarray}
\dot{\delta\rho} + 3H(\delta\rho + \delta P)
 + 3(\rho+P)\dot\R + a^{-1}\left[\nabla^2\delta q
+ (\rho+P) \nabla^2\left(a\dot{E}-{B}\right)\right] &=& 0\,,
\label{continuity}\\
\dot{\delta q} + 4H\delta q + a^{-1}\left[(\rho+P)A + \delta P +
{2\over 3} (\nabla^2+3K)\delta\pi\right]& =& 0 \,,
\label{momentum}
\end{eqnarray}
where the momentum perturbation is $\delta q = (\rho+P)(v+B)$.

The final 5-dimensional Einstein equation yields an equation of
state for the Weyl fluid~\cite{L2}, which in the 4-dimensional equations
follows from the symmetry properties of the projected Weyl tensor,
requiring $P_\E={1\over3}\rho_\E$ in the background and $\delta
P_\E={1\over3}\delta\rho_\E$ at first order.
We still require equations of motion for the effective energy and
momentum of the projected Weyl tensor, and these are provided by the
4-dimensional contracted Bianchi identities.
Note that these equations are {\em intrinsically four-dimensional,} only
being defined on the brane, and are not part of the five-dimensional
field equations.
The contracted Bianchi identities ($\nabla_\mu G^\mu_\nu=0$)
and energy-momentum conservation for matter on the brane
($\nabla_\mu T^\mu_\nu=0$) ensure, using Eq.~(\ref{modEinstein}), that
\begin{equation}
\nabla_\mu \E^\mu_\nu = \kappa_5^4\,\nabla_\mu\Pi^\mu_\nu \,.
\end{equation}
In the background we have
\begin{equation}
\dot\rho_\E +4H \rho_\E = 0 \,,
\end{equation}
and for the first-order perturbations we have
\begin{equation}
\label{Econtinuity} \dot{\delta\rho}_\E + 4H\delta\rho_\E +
4\rho_\E \dot\R +a^{-1}\left[\nabla^2\delta q_\E + {4\over3}
\rho_\E \nabla^2 \left(a\dot{E}-{B}\right)\right] = 0
\end{equation}
(which can be compared with the covariant form given
in~\cite{RoyI}). The key result here is that the effective energy
of the projected Weyl tensor is conserved independently of the
quadratic energy-momentum tensor. The only interaction is a
momentum transfer~\cite{SMS,RoyI}, as shown by the perturbed
momentum conservation equation
\begin{equation}
\dot{\delta q}_\E + 4H\delta q_\E +a^{-1}\left[ {4\over3}\rho_\E A
 + {1\over3} \delta \rho_\E + {2\over 3}
(\nabla^2+3K)\delta\pi_\E \right]= {(\rho+P )\over a\lambda}
\left[\delta\rho - 3Ha\delta q
  - (\nabla^2+3K) \delta\pi \right] \,,
\end{equation}
where the terms on the right hand side represent the momentum transfer
from the quadratic energy-momentum tensor. Note that the combination
$\delta\rho-3Ha\delta q$ that appears on the right hand side is
gauge-invariant and is equal to the density perturbation on
comoving hypersurfaces. Note also that, in the low-energy
regime, the right hand side becomes extremely small and one gets
a {\it quasi-conservation} for the Weyl momentum, in addition to the
exact conservation of the Weyl energy density.

\section{Curvature perturbations}

For the matter fluid we can define a gauge-invariant variable
corresponding to the curvature perturbation on hypersurfaces of
uniform density~\cite{Bardeen}
\begin{equation}
\zeta \equiv \R + {\delta\rho \over 3(\rho+P)} \,.
\label{zeta}
\end{equation}
The energy conservation equation~(\ref{continuity}) can then be
written as~\cite{WMLL}
\begin{equation}
\dot\zeta = - H \left({\delta P_{\rm nad} \over \rho+P}\right) -
{1\over3}
  \nabla^2\left({v\over a}+\dot{E}\right) \, ,
\label{dotzeta}
\end{equation}
where the non-adiabatic pressure perturbation is
\begin{equation}
\delta P_{\rm nad} \equiv \delta P - c_{\rm s}^2 \delta\rho \,.
\end{equation}
For adiabatic perturbations, $\zeta$ becomes a constant on large
scales, i.e., where gradient terms can be neglected~\footnote
{We assume that the universe looks locally like a FRW universe on
  sufficiently large scales. Thus the local momentum, shear and
  anisotropic stresses (and any other quantities derived from spatial
  gradients of scalars) must become negligible on large enough scales with
  respect to density, pressure or curvature perturbations.}.
This follows directly from the energy-conservation equation,
independently of the Einstein equations for the gravitational
field~\cite{WMLL}.
In fact it is possible to define a curvature perturbation, $\zeta_I$,
on hypersurfaces of uniform density for any matter component
separately, such as dust or radiation. This will also be constant for
adiabatic perturbations in that component on sufficiently large scales
if energy conservation holds for that component
separately~\cite{WMLL}, as shown in the Appendix.

We will also define a gauge-invariant curvature perturbation
on hypersurfaces of uniform total effective energy density on the
brane,
\begin{eqnarray}
\zeta_{\rm tot} &\equiv&
 \R + {\delta\rho_{\rm tot}\over 3(\rho_{\rm tot}+P_{\rm tot})}
\nonumber\\
&=& \R + {\delta\rho(1+\rho/\lambda) +
  \delta\rho_\E \over
  3(\rho+P)(1+\rho/\lambda)+4\rho_\E} \,.
\label{zetatot}
\end{eqnarray}

\subsection{Schwarzschild-Anti-de-Sitter background bulk}

If the projected Weyl tensor is non-vanishing in the background,
$\rho_\E\neq0$, then we can define a gauge-invariant curvature
perturbation on the brane
with respect to the projected Weyl tensor, entirely analogous to
that defined in Eq.~(\ref{zeta}) with respect to ordinary matter,
\begin{equation}
\zeta_\E = \R + {\delta\rho_\E \over 4\rho_\E} \,.
\end{equation}
Because the effective energy-momentum tensor of the projected Weyl
tensor has a definite equation of state and its effective energy is
conserved, the energy-conservation equation~(\ref{Econtinuity}) gives
\begin{equation}
\dot{\zeta}_\E=0\,,
\label{dotzetaE}
\end{equation}
on large scales (i.e., neglecting gradient terms).

The total curvature perturbation is then a weighted sum of $\zeta$ for
matter and $\zeta_\E$ for the projected Weyl tensor:
\begin{eqnarray}
\zeta_{\rm tot}& =& W \zeta + (1-W) \zeta_\E
\,,
\end{eqnarray}
where
\begin{equation}
\label{defW}
W = {3(\rho+P)(1+\rho/\lambda) \over
  3(\rho+P)(1+\rho/\lambda)+4\rho_\E} \,,
% W_\E &=& {4\rho_\E \over 3(\rho+P)(1+\rho/\lambda)+4\rho_\E} \,.
\end{equation}
It is instructive to study the behaviour of $W$. During the
low-energy radiation era, $W$ is a constant and $1-W \lesssim
\rho_\E/\rho$ is very small. In the subsequent matter era, $1-W$
will decrease like $a^{-1}$. %YYYY Finally,
By contrast, in a very high-energy radiation era ($\rho \gg \lambda$)
before the standard, i.e. low-energy, radiation era, $1-W$ would
increase like $a^4$. From this analysis, it is clear that $1-W$
reaches its maximum value during the low-energy radiation era, where
it is constrained by nucleosynthesis to be rather small
[$\lesssim0.03$ from Eqs.~(\ref{nuclimit}) and~(\ref{defW})]. During all
other eras, it will be still smaller.

If we define a (gauge-invariant) Weyl entropy perturbation,
\begin{equation}
{\cal S}_\E = \zeta_\E-\zeta
={\delta\rho_\E\over4\rho_\E}-{\delta\rho\over3(\rho+P)} \,,
\end{equation}
then we have
\begin{eqnarray}
\zeta_{\rm tot}&=&\zeta+(1-W){\cal S}_{\E} \,. \label{ztotSAdS}
\end{eqnarray}
On large scales,
\begin{eqnarray}
\dot{\zeta}_{\rm tot}
&=& W \dot\zeta + 3HW(1-W)\left( c_{\rm eff}^2 -
  {1\over3} \right) {\cal S}_\E\,.
\end{eqnarray}
Note that non-zero $\dot\zeta$ arises if there is a non-adiabatic
matter perturbation, whereas $\dot{\zeta}_\E=0$ always. Thus for
adiabatic matter perturbations, both $\zeta$ and $\zeta_\E$ are
constant, and the only change in $\zeta_{\rm tot}$ is then due to
the change in $W$ when ${\cal S}_\E\neq0$ and $c_{\rm eff}^2\neq
{1\over3}$.

{}From Eqs.~(\ref{dotzeta}) and (\ref{dotzetaE}), as long as we
may neglect gradient terms, we may express $\zeta_{\rm tot}$ as
\begin{eqnarray}
 \zeta_{\rm tot}(t)
=\zeta_*-W(t)\int_{t_*}^{t}dt'H\left({\delta P_{\rm nad}\over
\rho+P}\right) +\left[ 1-W(t) \right] {\cal S}_{\E*} \,,
\label{ztott}
\end{eqnarray}
where $t_*$ is some early epoch and $\zeta_*=\zeta(t_*)$, etc.
In particular, when the matter consists of
radiation and dust,
we have
\begin{eqnarray}
\zeta=\zeta_*+\left( {\rho_{\rm d}/\rho_{\rm r}\over4+3\rho_{\rm
    d}/\rho_{\rm r}}\right)S_{\rm dr}
\,, \label{zetadr}
\end{eqnarray}
where $\rho_{\rm r}$ and $\rho_{\rm d}$ are the radiation and dust
energy densities, respectively, and
\begin{equation}
 S_{\rm dr}
= 3\left( \zeta_{\rm d} - \zeta_{\rm r} \right)
= {\delta\rho_{\rm d}\over\rho_{\rm
 d}}-{3\over4}{\delta\rho_{\rm r}\over\rho_{\rm r}}\,,
\label{Sdr}
\end{equation}
is the entropy perturbation between the radiation and dust,
which remains constant on superhorizon scales
since $\zeta_{\rm d}$ and $\zeta_{\rm r}$ are separately conserved on
large scales.
The total curvature perturbation is then given by
\begin{eqnarray}
\zeta_{\rm tot} =\zeta_*+W\left({\rho_{\rm d}/\rho_{\rm
r}\over4+3\rho_{\rm d}/\rho_{\rm r}}\right)S_{\rm dr}  +
(1-W){\cal S}_{\E*} \,. \label{ztotrd}
\end{eqnarray}
The general form for $\zeta$ and $\zeta_{\rm tot}$ in a
multi-component matter system is given in the Appendix.

\subsection{Anti-de-Sitter background bulk}

If there is no projected Weyl tensor in the background, $\rho_\E=0$,
then any contribution from $\delta\rho_\E$ is non-adiabatic (and
automatically gauge-invariant).
The total curvature perturbation is then, on
using Eq.~(\ref{zeta}) in Eq.~(\ref{zetatot}),
\begin{equation}
\zeta_{\rm tot} = \zeta +
 {\delta\rho_\E \over 3(\rho+P)(1+\rho/\lambda)} \,.
\end{equation}
The continuity equation~(\ref{Econtinuity}) becomes
$\dot{\delta\rho}_\E + 4H\delta\rho_\E=0$ on large scales, and hence
\begin{equation}
\delta\rho_\E\propto {1\over a^{4}}\,.
\end{equation}
We find that
\begin{equation}
\dot\zeta_{\rm tot} = \dot\zeta + H \left( c_{\rm eff}^2 -
  {1\over3} \right) {\delta\rho_\E \over (\rho+P)(1+\rho/\lambda)} \,.
\end{equation}
%YYYY
The second term on the right may be compared with the expression
in~\cite{RoyII} for the total non-adiabatic pressure perturbation.
Note that in the very high-energy radiation era, this term is a
nonzero constant, whereas it is zero in the low-energy radiation
era.

Similar to Eq.~(\ref{ztott}), we may express $\zeta_{\rm tot}$ on
superhorizon scales in the present case as
%YYYY - H added to integral
\begin{eqnarray}
 \zeta_{\rm tot}=\zeta_*
-\int_{t_*}^{t}dt'H\left({\delta P_{\rm nad}\over \rho+P}\right)
+\left[{\delta\rho_{\E*}(a_*/a)^4\over
3(\rho+P)(1+\rho/\lambda)}\right] \,.
\end{eqnarray}
Then for a universe with radiation and dust, we have
\begin{eqnarray}
\zeta_{\rm tot}=\zeta_* +\left({\rho_{\rm d}/\rho_{\rm
r}\over4+3\rho_{\rm d}/\rho_{\rm r}}\right)S_{\rm dr} +
{1\over(4+3\rho_{d}/\rho_{\rm
r})(1+\rho/\lambda)}\,{\delta\rho_{\E*}\over\rho_{{\rm r}*}} \,.
\label{ztotAdSrd}
\end{eqnarray}

Before concluding this section, it is worthwhile to note the
following fact. If we introduce the initial entropy perturbation
$S_\E$ equivalent to ${\cal S}_{\E*}$ by
\begin{eqnarray}
  S_\E={\rho_\E\over\rho_{\rm r}}{\cal S}_{\E*}\,,
\end{eqnarray}
we can treat both $\rho_\E\neq0$ and $\rho_\E=0$ cases of the
radiation-dust universe in a unified manner.
Namely, Eqs.~(\ref{ztotrd}) and (\ref{ztotAdSrd}) are
expressed in the single form,
\begin{eqnarray}
\zeta_{\rm tot} = \zeta_* +\left[{\rho_{\rm d}/(\rho_{\rm
r}+\tilde\rho_\E)\over
    4+3\rho_{\rm d}/(\rho_{\rm r}+\tilde\rho_\E)}\right]S_{\rm dr}
+\left[ {4\over(4+3\rho_{\rm d}/\rho_{\rm
r})(1+\rho/\lambda)}\right]S_\E\,,
  \label{ztotunit}
\end{eqnarray}
where $\tilde\rho_\E=\rho_\E(1+\rho/\lambda)^{-1}$, and
$\zeta_*$, $S_{\rm dr}$ and $S_\E$ are constants to be determined
by the initial condition.

\section{Longitudinal gauge metric perturbations}

The curvature perturbation on hypersurfaces of uniform total
effective energy density, $\zeta_{\rm tot}$, can be directly
related, using the 4-dimensional (modified) Einstein equations, to
the gauge-invariant metric perturbations $\Phi$ and
$\Psi$~\cite{KS} in the longitudinal (or zero-shear or conformal
Newtonian) gauge:
\begin{eqnarray}
\zeta_{\rm tot} &=& \Phi - {3H\dot\Phi -3 H^2\Psi -
a^{-2}(\nabla^2+3K)\Phi \over 3(\dot{H}-Ka^{-2}) } \nonumber\\
{}&=& {(5+3w_{\rm tot})\over3(1+w_{\rm tot})}\,\Phi
+{2H\dot\Phi-(2/3a^2)(\nabla^2+6K)\Phi \over
3(H^2+Ka^{-2})(1+w_{\rm tot})} +{2a^2H^2 \over \rho_{\rm
tot}(1+w_{\rm tot})}\,\delta\pi_{\rm tot} \,,\label{zetatot2}
\end{eqnarray}
where $w_{\rm tot}=P_{\rm tot}/\rho_{\rm tot}$ and
\begin{eqnarray}
&&\Psi= A+\left[a(B-a\dot E)\right]\!\dot{\vphantom{E}}\,,
% \label{}
\\ &&\Phi= \R +\dot a(B-a\dot E)\,, 
% \label{}
\\
&&\delta\pi_{\rm
tot}=\left(1-{\rho+3P\over2\lambda}\right)\delta\pi
+\delta\pi_\E\,.
\end{eqnarray}

Equation~(\ref{zetatot2})
is obtained by using the background (modified) Friedmann
equations, i.e., the standard ones with `total' fluid as matter, and the
traceless part of the spatial perturbed (modified)
Einstein equations, which yields
\begin{equation}
\label{traceless}
\Phi+\Psi=-\kappa_4^2 a^2\delta\pi_{\rm tot}\,,
\end{equation}
as in general relativity~\cite{KS}.
Note that in the absence of anistropic stresses ($\delta\pi_{\rm
  tot}=0$) there is essentially only one gauge-invariant scalar metric
perturbation, $\Phi=-\Psi$, which is determined directly on large
scales from the primordial $\zeta_{\rm tot}$. But in the presence of
anisotropic stresses it will no longer be possible to determine the
metric perturbations from $\zeta_{\rm tot}$ alone. Even if there are
  no (or negligible) matter anisotropic stresses, $\delta\pi_{\rm
  tot}=\delta\pi_\E$ may be non-zero so that $\Phi+\Psi\neq0$.

In this section, our goal will be to relate the curvature
perturbations to the metric perturbations, $\Phi$ and $\Psi$,
using the results obtained in the previous section for the
curvature perturbations. This will be useful in the next section
where we will compute the large-scale anisotropies.  This is also
useful if one wishes to make the connection between the primordial
curvature fluctuations and the late-time metric fluctuations.  For
simplicity, we will assume here that the universe is spatially
flat.  We will distinguish the three following cases of
cosmological interest: a (low-energy) dust dominated era, a
(low-energy) radiation dominated era, and finally a very
high-energy radiation era.

\subsection{Dust dominated era}

In a dust-dominated universe, the curvature perturbation
$\zeta_{\rm tot}$ given by Eq.~(\ref{ztotunit}) reduces to
\begin{eqnarray}
 \zeta_{\rm tot}=\zeta_*+{1\over3}S_{\rm dr}+{4\rho_{\rm r}\over3\rho_{\rm
 d}}S_\E\,.
 \label{ztotdust}
\end{eqnarray}
The origin of each term on the right hand side is apparent.
The first describes the adiabatic perturbation, the second
 the primordially isocurvature perturbation, and the third
the Weyl entropy perturbation.
On the other hand, Eq.~(\ref{zetatot2}),
for a spatially flat ($K=0$) FRW cosmology on large scales,
reduces to
\begin{eqnarray}
 \zeta_{\rm tot}={5\over3}\Phi+{2\over3}a{d\over da}\Phi
+{2\kappa_4^2\over 3}a^2\delta\pi_{\rm tot}\,.
\end{eqnarray}
Hence we find that the parts of $\Phi$ corresponding to each term in
Eq.~(\ref{ztotdust}) at the dust-dominated stage are given by
\begin{eqnarray}
  \Phi_{\rm ad}={3\over5}\zeta_*\,,\quad
  \Phi_{\rm iso}={1\over5}S_{\rm dr}\,,\quad
  \Phi_\E=\left({4\rho_{\rm r}\over3\rho_{\rm d}}\right)S_\E\,,
\quad \Phi_\pi=-{\kappa_4^2\over a^{5/2}}\int %YYYY ^a
\delta\pi_{\rm tot}\,a^{7/2}da\,,
  \label{Phidust}
\end{eqnarray}
where $\Phi_\pi$ is the part due to the anisotropic stress
perturbation which does not contribute to $\zeta_{\rm tot}$ on large
scales. The corresponding parts of $\Psi$ are readily calculated
from Eq.~(\ref{traceless}) as
\begin{eqnarray}
  \Psi_{\rm ad}=-{3\over5}\zeta_*\,,\quad
  \Psi_{\rm iso}=-{1\over5}S_{\rm dr}\,,\quad
\Psi_\E=-\left( {4\rho_{\rm r}\over3\rho_{\rm
d}}\right)S_\E\,,\quad
\Psi_\pi={\kappa_4^2\over a^{5/2}}\int %YYYY^a
\delta\pi_{\rm tot}\,a^{7/2}da
 -\kappa_4^2 a^2\delta\pi_{\rm tot}\,.
  \label{Psidust}
\end{eqnarray}

\subsection{Low-energy radiation era}

We repeat here the same analysis as before for the (low-energy)
radiation era
with $\lambda\gg\rho_{\rm r}\gg\rho_{\rm d}$.
In this case the curvature perturbation $\zeta_{\rm
  tot}$ given by Eq.~(\ref{ztotunit}) simply reduces to
\begin{eqnarray}
 \zeta_{\rm tot}=\zeta_*
 +{\rho_{\rm d}\over 4(\rho_{\rm r}+\rho_\E)}S_{\rm dr}
 + S_\E\,.
 \label{ztotrad}
\end{eqnarray}
Here $w_{\rm tot}={1\over3}$, and the relation between the total
curvature perturbation and the metric fluctuation, given by
Eq.~(\ref{zetatot2}), reduces to
\begin{eqnarray}
 \zeta_{\rm tot}={3\over 2}\Phi+{1\over 2}a{d\over da}\Phi
+{\kappa_4^2\over 2}a^2\delta\pi_{\rm tot}\,.
\end{eqnarray}
One thus gets
\begin{eqnarray}
 \Phi_{\rm ad}={2\over 3}\zeta_*\,,\quad
\Phi_{\rm iso}={1\over8}\left({\rho_{\rm d}\over \rho_{\rm
r}+\rho_\E}\right)S_{\rm dr} \,,\quad
  \Phi_\E={2\over3}S_\E\,,\quad
 \Phi_\pi=-{\kappa_4^2\over a^{3}}\int %YYYY ^a
\delta\pi_{\rm tot}\,a^{4}da\,.
  \label{Phirad}
\end{eqnarray}

\subsection{Very high-energy radiation era}

We finally consider the case of a radiation era where the
background evolution is highly non-standard, with $\rho_{\rm r}\gg
\lambda$, as well as $\rho_{\rm r}\gg \rho_{\rm d}$. Equation
(\ref{ztotunit}) gives
\begin{eqnarray}
 \zeta_{\rm tot}=\zeta_*
+ {\rho_{\rm d}\over 4\rho_{\rm r}} S_{\rm dr}
 + {\lambda\over \rho} S_\E\,,
 \label{ztotradhe}
\end{eqnarray}
and, as in the dust-dominated case, the contribution from the Weyl
component in $\zeta_{\rm tot}$ is time-dependent.
Since $w_{\rm tot}={5\over3}$, Eq.~(\ref{zetatot2}) now yields,
\begin{eqnarray}
 \zeta_{\rm tot}={5\over 4}\Phi+{1\over 4}a{d\over da}\Phi
+{\kappa_4^2\over 2}a^2\delta\pi_{\rm tot}\,.
\end{eqnarray}
One then gets
\begin{eqnarray}
\Phi_{\rm ad}={4\over 5}\zeta_*\,,\quad \Phi_{\rm
iso}={1\over6}\left({\rho_{\rm d}\over \rho_{\rm r}}\right) S_{\rm
dr} \,,\quad
  \Phi_\E={4\over 9}\left({\lambda\over \rho}\right)S_\E\,,\quad
 \Phi_\pi=-{\kappa_4^2\over a^{5}}\int %YYYY ^a
\delta\pi_{\rm tot}\,a^{6}da\,.
  \label{Phiradhe}
\end{eqnarray}

\section{Large-angle CMB anisotropy}

Let us consider the large-angle CMB anisotropy in our scenario.
Since the energy density $\rho$ is much smaller than the
tension $\lambda$ at and after the decoupling of photons
and baryons, we may safely neglect the $\rho/\lambda$
corrections in all the equations.

Assuming a spatially flat universe,
the (generalized) Sachs-Wolfe effect is described as\cite{KS2}
\begin{eqnarray}
 \left({\delta T\over T}\right)_{\sc sw}(\vec{\gamma},\eta_0)=
\left({1\over4}\Delta_{\rm s,r} +\Psi\right)(\eta_{\rm
dec},\vec{x}(\eta_{\rm dec})) +\int_{\eta_{\rm dec}}^{\eta_0}d\eta
\,\partial_{\eta}\left(\Psi-\Phi\right)(\eta,\vec{x}(\eta)),
\label{sachswolfe}\end{eqnarray} where
$\vec{x}(\eta)=\vec{\gamma}(\eta_0-\eta)$, $\eta$ is the conformal
time ($d\eta=dt/a(t)$), and $\Delta_{\rm s,r}=\delta\rho_{\rm
r}/\rho_{\rm r}+4Ha(a\dot E-B)$ is the photon density perturbation
on the shear-free hypersurfaces.\footnote{For a spatially curved
universe, the only change in the Sachs-Wolfe formula is the
expression for $\vec{x}(\eta)$, which can only be obtained by
integrating the null geodesic equations.} The last integral along
photon null geodesics is called the integrated Sachs-Wolfe effect.
In contrast with it, for convenience, let us call the first two
terms in the parentheses the `direct' Sachs-Wolfe effect.

By evaluating the right hand side of Eq.~(\ref{zeta}) in the
shear-free gauge, we find the curvature perturbation on
hypersurfaces of uniform photon density, $\zeta_{\rm r}$, is
expressed in terms of $\Delta_{\rm s,r}$ and $\Phi$ as
\begin{eqnarray}
 \zeta_{\rm r}=\Phi+{1\over4}\Delta_{\rm s,r}\,.
\label{zetar}
\end{eqnarray}
Thus the Sachs-Wolfe formula (\ref{sachswolfe}) may be expressed
as
\begin{eqnarray}
 \left({\delta T\over T}\right)_{\sc sw}(\vec{\gamma},\eta_0)=
\left(\zeta_{\rm r}+\Psi-\Phi\right)(\eta_{\rm
dec},\vec{x}(\eta_{\rm dec})) +\int_{\eta_{\rm dec}}^{\eta_0}d\eta
\,\partial_{\eta}\left(\Psi-\Phi\right)(\eta,\vec{x}(\eta)).
\label{sachswolfe2}
\end{eqnarray}

To evaluate the quantities appearing in the Sachs-Wolfe formula,
let us further assume the universe is dust-dominated at decoupling.
One can thus apply the results obtained in the previous section in the
case of a  dust-dominated universe. Moreover,  $\zeta_{\rm r}$ is
related to $\zeta$ as
\begin{eqnarray}
 \zeta_{\rm r}
% =\zeta-(\zeta-\zeta_{\rm r})
=\zeta-\left({\rho_{\rm d}\over3\rho_{\rm d}+4\rho_{\rm r}}\right)
S_{\rm dr}\,. \label{zetardr}
\end{eqnarray}
Comparing this with Eq.~(\ref{zetadr}), we find
\begin{equation}
 \zeta_{\rm r}=\zeta_*\,.
\label{zetarsol}
\end{equation}
Thus the curvature perturbation on hypersurfaces of uniform photon
density exactly represents the adiabatic curvature perturbation.

Gathering all the terms given in
Eqs.~(\ref{Phidust}), (\ref{Psidust}) and (\ref{zetarsol}) together,
the terms contributing to the direct Sachs-Wolfe effect become
\begin{eqnarray}
\label{directSW} \zeta_{\rm r}+\Psi-\Phi
=-{1\over5}\zeta_*-{2\over5}S_{\rm dr} -{8\over3}\left({\rho_{\rm
r}\over\rho_{\rm d}}\right)S_\E -\kappa_4^2 a^2\delta\pi_{\rm tot}
+{2\kappa_4^2\over a^{5/2}}\int %YYYY _0^a
\delta\pi_{\rm tot}\,a^{7/2}da\,.
\end{eqnarray}
The first and second terms on the right hand side describe the
conventional adiabatic and isocurvature Sachs-Wolfe effects, which
may be expressed in terms of $\Psi$ as ${1\over3}\Psi_{\rm ad}$
and $2\Psi_{\rm iso}$, respectively. The third term due to the
Weyl entropy perturbation may be expressed as $2\Psi_\E$. One may
be tempted to regard it as a kind of isocurvature perturbation.
However, if we recall Eq.~(\ref{ztotunit}), we see it gives a
time-independent contribution to $\zeta_{\rm tot}$ during the
radiation-dominated stage. The magnitude of $S_\E$ depends very
much on the early history of the universe. If the universe
undergoes inflation at an early stage, $S_\E$ will be totally
negligible after inflation. Observationally the strongest
constraint on $S_\E$ comes from the COBE CMB
anisotropies\cite{COBE};
\begin{equation}
S_\E\lesssim 10^{-4} \,,
\end{equation}
since $(\rho_{\rm r}/\rho_{\rm d})\sim 0.1$ at decoupling.

Note that one can also relate the perturbations at last scattering
to the perturbations in the early universe where these perturbations
might have been generated. Consider for instance the adiabatic part of
the metric perturbations. $\Psi_{\rm ad}$ at last scattering, i.e.
during dust domination, is related to the corresponding primordial
perturbation in a low -energy radiation era by
\begin{equation}
\Psi_{\rm ad}= {9\over 10}\Psi_{\rm ad,r(low)},
\end{equation}
and to the primordial perturbation in a very high-energy radiation era
by
 \begin{equation}
\Psi_{\rm ad}= {3\over 4}\Psi_{\rm ad,r(high)}.
\end{equation}

The last two terms in Eq.~(\ref{directSW}) due to anisotropic stress
are generally negligible except for
possibly the Weyl contribution, $\delta\pi_\E$, which cannot be
theoretically constrained within the present approach.
Even inflation does not seem to be necessarily
effective for reducing the amplitude of $\delta\pi_\E$.
The magnitude of anisotropic stress on a comoving scale $k^{-1}$ is
given by
\begin{eqnarray}
 {|\delta\pi_{ij}|\over\rho+P}\sim
{k^2\over a^2H^2}\left({a^2H^2\delta\pi\over \rho+P}\right)\,,
\end{eqnarray}
where $k$ is the comoving wavenumber. Assuming all the
perturbations behave regularly in the limit $t\to0$ and in the
large-scale limit, the only restriction is that
$H^2a^2\delta\pi_{\E}/(\rho+P)$ be regular in both of the limits.
{}From the amplitude of the COBE CMB anisotropies\cite{COBE} and
Eq.~(\ref{directSW}), we obtain the observational bound
\begin{equation}
\kappa_4^2 a^2 \delta\pi_\E \lesssim 10^{-5} \,.
\end{equation}

Finally, we consider the integrated Sachs-Wolfe effect. In
addition to the conventional contributions discussed in the
literature~\cite{HuSugi}, there are contributions specific to our
scenario. From Eqs.~(\ref{Phidust}) and (\ref{Psidust}), we have
\begin{eqnarray}
 \partial_\eta(\Psi-\Phi)
=-\partial_\eta\left({8\over3}{\rho_{\rm r}\over\rho_{\rm d}}S_\E
+\kappa_4^2a^2\delta\pi_{\rm tot} -{2\kappa_4^2\over
a^{5/2}}\int %YYYY _0^a
\delta\pi_{\rm tot}\,a^{7 /2}da\right)\,.
\end{eqnarray}
The first term due to $S_\E$ gives the same effect as the one due to
insufficient dust-dominance in conventional 4-dimensional cosmological
models\cite{HuSugi}. It is effective only in the vicinity of the last
scattering surface, and its effect is expected to be of the same order
of magnitude as the corresponding contribution in the direct
Sachs-Wolfe effect.  On the other hand, we are unable to
constrain the magnitude of the last two terms due to $\delta\pi_{\rm
tot}$, because of the presence of the Weyl anisotropic stress,
$\delta\pi_\E$, whose behavior is undetermined within our approach.

\section{Discussion}

In this paper we have shown that it is possible to extend some results
for the evolution of scalar perturbations about four-dimensional FRW
cosmological solutions to a five-dimensional brane-world scenario by
working solely with the induced four-dimensional Einstein equations on
the brane.  In particular, the curvature perturbation on uniform
matter density hypersurfaces, $\zeta$, remains constant for adiabatic
matter perturbations on sufficiently large scales, where gradient
terms become negligible.  This remains applicable in a wide variety of
higher-dimensional models so long as local conservation of energy
holds for some or all matter fields on the four-dimensional
brane-world.

We have focused on the case of five-dimensional Einstein gravity with
a cosmological constant in the bulk, which ensures energy-conservation
for matter on the brane.  In addition to ordinary cosmological matter,
a new component appears in the induced four-dimensional Einstein
equations on the brane, which is the manifestation of the
five-dimensional bulk gravitons. This component, which we call the
Weyl component because it corresponds to the projected
five-dimensional Weyl tensor, can be described effectively as a fluid
from the brane point of view, which may appear in the background
solution, but is constrained to remain small with respect to the
ordinary radiation component.

The effective energy of the Weyl component is locally conserved
independently of ordinary matter for linear perturbations, even though
there may be momentum transfer at high energies. We are therefore able
to define another perturbation, $\zeta_\E$, when $\rho_\E\neq0$,
corresponding to the curvature perturbation on hypersurfaces of
uniform effective Weyl density, which remains constant on large
scales. If $\rho_\E=0$, then $\delta\rho_\E$ is a (gauge-invariant)
non-adiabatic perturbation whose evolution is determined by the energy
conservation equation. We are then able to model the evolution of the
total effective curvature perturbation for matter plus Weyl fluid,
$\zeta_{\rm tot}$, constructed from the matter $\zeta$ and the Weyl
fluid $\zeta_\E$ or $\delta\rho_\E$. This in turn can be related to
the longitudinal-gauge metric perturbation, $\Phi$, either in the
early radiation-dominated era (where non-conventional background
evolution can change the usual relation between $\Phi$ and $\zeta_{\rm
  tot}$ at very high energies), or later during dust-domination.

We have also studied the possible impact upon cosmic microwave
background anisotropies. The presence of the Weyl component has
essentially two possible effects. On the one hand, there is an
additional contribution from the Weyl entropy perturbation $S_\E$ that
is similar to an extra isocurvature contribution. On the other hand,
the anisotropic stress of the Weyl component, $\delta\pi_\E$, also
contributes to the CMB anisotropies.  In the absence of anisotropic
stresses, the curvature perturbation $\zeta_{\rm tot}$ is sufficient
to determine the metric perturbation $\Phi$ and hence the large-angle
CMB anisotropies. However bulk gravitons can also generate anisotropic
stresses which, although they do not affect the large-scale curvature
perturbation $\zeta_{\rm tot}$, can affect the relation between
$\zeta_{\rm tot}$ and $\Phi$ and hence the CMB anisotropies on large
angles.
There is no intrinsic brane equation determining the evolution of
$\delta\pi_\E$ and thus allowing us to estimate it during and after
inflation. On intuitive grounds, the part of $\delta\pi_\E$ generated
by density inhomogeneity on the brane is expected to be no greater
than the matter anisotropic stress $\delta\pi$, which may be neglected
for calculating large-angle CMB anisotropies.  As for the other part
due to quantum fluctuations of scalar gravitons during inflation, a
dimensional analysis suggests $\kappa_4^2a^2\delta\pi$ is at most of
the order of $\kappa_4^2H^2$.  However, this remains to be proved.

Therefore, while the present approach based on the study of the
perturbation equations solely on the brane has led us to significant
results on large scales, it has also clearly shown us its limits.
There is still a need to determine the evolution of the metric
perturbations in the bulk in order to determine (i) the amplitude of
the Weyl anisotropic stress $\delta\pi_\E$, and (ii) the evolution of
metric perturbations on the brane at sub-Hubble wavelengths.  In this
respect, it is important to devise a specific model for the
description of the bulk metric perturbations.  One would then have to
relate the bulk perturbations with their fluid description on the
brane in terms of an energy density, a momentum density and
anisotropic stress \cite{L2,DDK}.  Some such modelling of the
evolution of perturbations will be required to make predictions for
the shape of the CMB power spectrum over a range of angular scales, to
compare with existing observational data.

\acknowledgments

DW is supported by the Royal Society. MS is supported in part by
Yamada Science Foundation, and by PPARC
for a visit to Portsmouth, during which this work was initiated.
MS would like to thank all the members of Relativity and Cosmology Group,
University of Portsmouth, for warm hospitality.

\appendix
\section{Extension to multi-component matter}
Here we present an extension of our approach to a multi-component
matter system.
The curvature perturbation on hypersurfaces of uniform
$I$-th matter density is defined by
\begin{eqnarray}
 \zeta_I={\cal R}+{\delta\rho_I\over3(\rho_I+P_I)}\,.
\end{eqnarray}
For simplicity, let us assume $\rho_\E\neq0$.
Then defining the weight $W_I$ for the $I$-th component by
\begin{eqnarray}
 &&W_I={3(\rho_I+P_I)\over3(\rho+P)+4\tilde\rho_\E}\quad\mbox{for}\quad
 I\neq\E\,, \nonumber\\
 &&W_\E={4\tilde\rho_\E\over3(\rho+P)+4\tilde\rho_\E}\,,
\end{eqnarray}
where $\tilde\rho_\E=\rho_\E(1+\rho/\lambda)^{-1}$,
we have
\begin{eqnarray}
 \zeta_{\rm tot}=\sum_{I}W_I\zeta_I\,.
\end{eqnarray}
If $\rho_\E=0$, the only modification is to
replace the Weyl contribution as
\begin{eqnarray}
 W_\E\zeta_\E\quad\to\quad
 {\delta\rho_\E\over3(\rho+P)(1+\rho/\lambda)+4\rho_\E}\,.
\end{eqnarray}

If we assume all the components are non-interacting
with each other, then the energy-momentum conservation equations
(\ref{continuity}) and (\ref{momentum}) hold for each component separately.
{}From the energy conservation (\ref{continuity}), the equation of
motion for $\zeta_I$ is given by
\begin{eqnarray}
\dot\zeta_I=-H\left({\delta P_{I,\rm nad}\over\rho_I+P_I}\right)
-{1\over3}\nabla^2\left({v_I\over a}+\dot E\right)\,,
\end{eqnarray}
where $v_I$ is the velocity potential of the $I$-th matter component.
Thus on sufficiently large scales,
\begin{eqnarray}
\dot\zeta_I=-H\left({\delta P_{I,\rm
nad}\over\rho_I+P_I}\right)\,,
\end{eqnarray}
and $\zeta_I$ will remain constant on large scales for adiabatic
perturbations of the $I$-th matter component, such that $\delta
P_{I,\rm nad}=0$.

\end{document}